\documentclass[aps,prl,twocolumn,showpacs]{revtex4}

\usepackage{epsfig}
\usepackage{subfigure}

\usepackage{times}

\newcounter{lastnote}

\def\e{\begin{equation}}
\def\f{\end{equation}}
\def\_#1{{\bf #1}}

\def\.{\cdot}

\def\E{\epsilon}

\def\l#1{\label{eq:#1}}
\def\r#1{(\ref{eq:#1})}
\def\=#1{\overline{\overline #1}}
\def\##1{{\bf#1\mit}}

\begin{document}

\title{An ultra-broadband electromagnetically indefinite medium formed by aligned carbon nanotubes}


\author{Igor Nefedov, Sergei Tretyakov, Constantin Simovski}
\affiliation{Aalto University,
School of Electrical Engineering  \\
SMARAD Center of Excellence,
P.O. Box 13000, 00076 Aalto, Finland}


\date{}


\begin{abstract}%

Anisotropic materials with different signs of components of the permittivity tensor are called {\itshape indefinite materials}. Known realizations of indefinite media suffer of high absorption losses. We show that  periodic arrays of parallel carbon nanotubes (CNTs) can behave as a low-loss indefinite medium in the infrared range. We show that a finite-thickness slab of CNTs supports the propagation of backward waves with small attenuation in an ultra-broad frequency band. In prospective, CNT arrays can be used for subwavelength focusing and detection, enhancing the radiation efficiency of small sources.
\end{abstract}


\maketitle

Metal-dielectric
nanostructures \cite{Ritchie,Kretschmann} have already brought applications to microscopy and sensing
\cite{Hecht} and have the potential for realization of future
nanosized optical devices and circuits \cite{Engheta}.
In 2003 the authors of \cite{Smith} noticed the potential of
so-called {\itshape indefinite metamaterials} for subwavelength
imaging of objects at electrically large distances from them (the
concept of such imaging was first introduced in \cite{Pendry}).
Indefinite metamaterials are artificial uniaxial materials in which
the axial and tangential components of the permittivity and
permeability tensors have different signs. In these materials the
isofrequency surfaces
have hyperbolic shape. This results
in a possibility to design a ``hyperlens'', where evanescent near
fields are transformed into propagating modes and can be transported
at electrically long distances \cite{Zhang}.

The main challenge on the way to realize this and other effects is
in the realization of low-loss indefinite materials.
For waves of only one
linear polarization (TM-polarized waves with respect to the axis of
positive permittivity) it is enough to realize a layer of an
indefinite \emph{dielectric} metamaterial \cite{Smith3}, whose
permeability is unity and only components of the permittivity tensor
have different signs. For the visible range such a metamaterial was
designed in \cite{Smith3} as an array of parallel plasmonic (metal) nanowires.
Since all plasmonic phenomena are
related to strong dissipation, the axial component of the
permittivity has a significant imaginary part which
strongly restricts the subwavelength imaging property of the
hyperlens \cite{Smith3}.
In the microwave range, materials with negative permittivity can be
realized as arrays of thin metal wires \cite{Brown1,Brown2}.
However, if the field varies along the wires, the properties of the
structure are more complicated that those of a continuous medium due
to strong spatial dispersion \cite{SpatD}.
The spatial dispersion can be suppressed partially \cite{Hudl}
or even totally \cite{Demetr,Grib,stas}. However,
manufacturing of structures \cite{Hudl,Demetr,Grib,stas} becomes a challenge already at
millimeter waves as dimensions need to be quite small for
high-frequency applications.
For electromagnetic waves in the THz and mid infrared (MIR) range
there are
no known structures with an indefinite permittivity tensor. The
imaginary and real parts of complex permittivity of metals in this
range are of the same order and rather high.

Here we show that arrays of metallic carbon nanotubes (CNT) behave
as ultra wide-band and low-loss indefinite materials in the THz and MIR
ranges. This collective property of arrays results from the
quantum properties of individual  CNT, namely their high quantum
inductance, so-called kinetic inductance \cite{Lutt}, which leads to
suppression of spatial dispersion in arrays of parallel CNTs. This
can open a possibility for realization of interesting
devices, which we discuss here.

Two-dimensional  periodic arrays of carbon nanotubes are fabricated
by many research groups. Such CNT arrays form finite-thickness slabs
which are used already as field emitters \cite{ShFan}, biosensors
\cite{YLin} and antennas \cite{Wang,DresselNat}.
We assume that all
nanotubes possess metallic properties that is a realistic
assumption in view of recent studies of single-wall CNTs \cite{WonBong,Kusunoki,modified}.
Usually
carbon nanotubes form hexagonal lattices in arrays in processes of
fabrication. However, for low-density arrays for which
arrangement of nanotubes is not important. By this reason and for
simplicity we assume lattices to be square with the constant $d$.

For eigenwaves propagating in arrays of infinitely long carbon nanotubes we take the space-time dependence of fields and currents as $\exp{[j(\omega t-k_zz)]}$.
Effective electromagnetic properties of CNT arrays can be
understood from  the nonlocal quasistatic model of wire media (WM)
\cite{stas}. In the framework of the effective medium model the CNT
array can be considered as a uniaxial material with the permittivity
dyadic \e\l{p1} \=\E=\E_{zz}\_z_0\_z_0+\E_0(\_x_0\_x_0+\_y_0\_y_0),
\f where $\E_0$ is the permittivity of vacuum (we consider CNTs
placed in vacuum). As was shown in \cite{stas},
\e\l{p2}\begin{array}{cc}
\frac{\E_{zz}}{\E_0}=1-\frac{k_p^2}{k^2-j\xi k-k_z^2/n^2}, &
k_p^2=\frac{\mu_0}{d^2L_{\rm cnt}},
\end{array}
 \f
where $k_p$ is
the effective plasma wavenumber $n^2=L_{\rm cnt}C_{\rm cnt}/(\E_0\mu_0)$, $L_{\rm cnt}$, $C_{\rm cnt}$ are the effective inductance and capacitance
per unit length, respectively, and the parameter $\xi=(R_{\rm cnt}/L_{\rm cnt})\sqrt{\E_0\mu_0}$ is responsible for losses.
Distributed parameters $L_{\rm cnt}$, $C_{\rm cnt}$ and $R_{\rm cnt}$ for a separate carbon nanotube can be obtained using the model of impedance cylinder and effective boundary conditions developed in \cite{Slepyan}.
According to this model the simple approximate expression for the complex surface conductivity, which is valid for metallic zigzag CNTs at frequencies below optical transitions looks as:
\e\l{a1} \sigma_{zz}\cong
-j\frac{2\sqrt{3}e^2\Gamma_0}{3q\pi\hbar^2(\omega-j\nu)}. \f
Here
$e$ is the electron charge, $\Gamma_0=2.7$~eV is the overlapping
integral, $\tau=1/\nu$ is the relaxation time, $\hbar$ is the
reduced Planck constant. The radius of metallic zigzag CNT, taken as an example, $r$ is determined
by an integer $q$ as $r=3\sqrt{3}qb/2\pi$, where $b=0.142$\,nm is
the interatomic distance in graphene.
Since the wall of a
single-wall CNT is a monoatomic sheet of carbon, \r{a1} can be
considered as the surface conductivity of the carbon nanotube. The
surface impedance per unit length
\e\l{z}
z_i=\frac{1}{2\pi
r\sigma_{zz}}=\frac{\sqrt{3}q\hbar^2\nu}{4e^2\Gamma_0r}+
j\omega\frac{\sqrt{3}q\hbar^2}{4e^2\Gamma_0r}=R_0+j\omega L_0. \f

Let us find $L_{\rm cnt}$, $C_{\rm cnt}$ and $R_{\rm cnt}$ for a separate carbon nanotube.
The total inductance per unit length
$L_{\rm cnt}=L_0+L_{\rm em}$, where $L_0$ is the kinetic inductance  \cite{Lutt}
defined by the formula \r{z}, which has a quantum nature, and $L_{\rm em}$ is the electromagnetic
inductance per unit length.
For thin CNTs the
kinetic inductance strongly dominates over the electromagnetic inductance,
e.g., for the zigzag CNT,
having
the radius $r\simeq 1.53$\,nm ($q=13$), $L_{\rm em}=5.6\times 10^{-7}$\,H/m
and $L_0=3.7\times10^{-3}$\,H/m. This shows that the electromagnetic
inductance can be neglected.
The total capacitance
of CNTs $C_{\rm cnt}$ includes the electrostatic capacitance defined
as $C_s=\E_0\mu_0/L_{\rm em}$ and the quantum capacitance
$C_q=2e^2/(hv_F)$, where $v_F$ is the Fermi velocity which is equal
to $8\times10^5$\,F/m for graphene and CNT \cite{BurkeNT}. The two effective capacitances are connected
serially. For the considered example $C_s\simeq 1.98\times
10^{-11}$\,F/m, $C_q\simeq 9.66\times 10^{-11}$\,F/m and $C_{\rm
tot}=C_eC_q/(C_e+C_q)=1.64\times 10^{-11}$\,F/m.
Let us estimate the terms entering formula \r{p2} at the
frequency $\omega/2\pi=1$\,THz.
One obtains $k_p^2=4.32\times
10^{10}$\, 1/m$^2$, $k^2=4.39\times 10^8$\,1/m$^2$ and dimensionless
$n^2=4.32\times 10^7$, so the last term in the denominator of formula \r{p2} can be neglected.

\begin{figure}
 \centering \epsfig{file=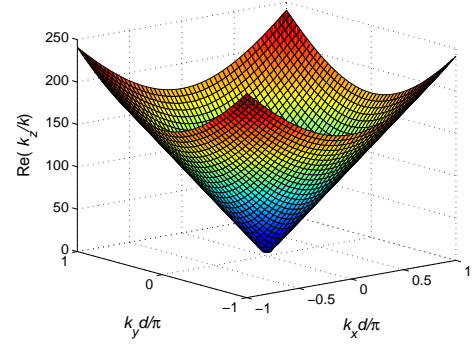, width=6.5cm}
\caption{(color online). The real part of the slow-wave factor in the transversal plane, calculated at $\omega/2\pi=1$\,THz for $d=2.5$\,nm.}
 \label{eff}\end{figure}

Thus
strong spatial dispersion is suppressed and the
material behaves as a uniaxial free-electron plasma, where electrons can move only
along $z$-direction.
Apparently, this corresponds to an indefinite medium at frequencies below the
plasma frequency, because the permittivity in the directions
orthogonal to the tubes is close to unity.
Dispersion equation for waves propagating in a uniaxial crystal is \cite{Landau}
\e\l{y1}
\E_0k_{\perp}^2=\E_{zz}(k^2-k_z^2),\f
where $k_{\perp}^2=k_x^2+k_y^2$.
Substituting \r{p2} with $k_z^2/n^2=0$ into Eq.~\r{y1} we obtain a simple formula:
\e\l{p7}
k_z^2=\frac{k^2(k^2-k_{\perp}^2-k_p^2)}{k^2-k_p^2}\f
which describes a typical conic-type dispersion, see Fig.~\ref{eff}.
Applicability of the described effective medium model  was verified
comparing with the results of the electrodynamic model, based on Green's
function method \cite{IgorPRB}. This model takes into account the array periodicity and demonstrates that the effective medium model gives a good agreement with the more accurate theory for small transversal wave numbers ($|k_{\perp}d|<0.15\pi$).

In a series of papers published in 2003-2004 by the groups of G.
Eleftheriades,  C. Caloz and T. Itoh, Pendry's concept of
perfect lens \cite{Pendry} was experimentally confirmed for a planar
analogue of the perfect lens (see also books
\cite{Elefther} and \cite{Caloz1}).
Planar networks composed of backward-wave transmission lines (TLN)
are surface analogues of bulk double-negative materials.
Surface formed by TLN acts as a layer of an
indefinite medium, where the spatial dispersion is suppressed
 \cite{Olli}. Based on the results of the previous duscussion, we expect
that in the THz and MIR ranges a similar 
backward-wave structure can be realized as an array of aligned CNTs.

Let us consider waves propagating in a finite-thickness slab of CNTs
assuming for simplicity that the slab is placed between perfect
electric conductor (PEC) and perfect magnetic conductor (PMC)
planes, where the PMC boundary models the open-ended interface with free space.
The resonant condition for the longitudinal wavenumber $k_z$ reads  $k_z=\pi/(2h)$ where $h$ is
the thickness of the slab. The relation between the transversal wave
vector components and the wavenumber in free space is the following:
\e\l{p8} k_{\perp}^2=k_x^2+k_y^2=\frac{(k^2-k_p^2)(k^2-k_z^2)}{k^2}.\f
One can easily show that the
derivative ${\rm d}k_{\perp}^2/{\rm d}k^2<0$ if $k_z/k>1$ and $k_p/k>1$.
Thus, in this regime the finite-thickness slab supports propagation of backward
waves, because the group velocity is in the opposite direction to
the phase velocity.
\begin{figure}
\centering \epsfig{file=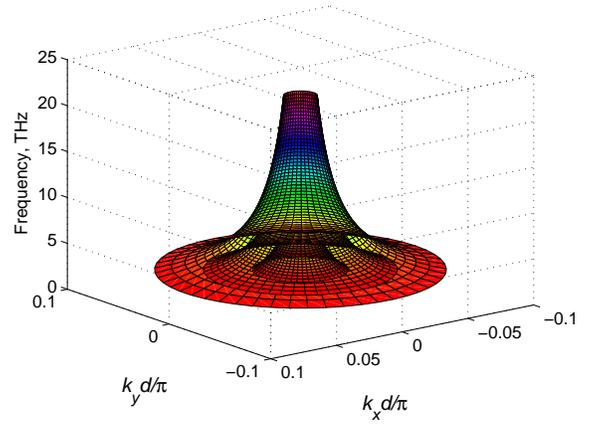, width=8cm}
\caption{(color online).
Frequency as a function of the normalized wavenumbers in the transverse plane $k_xd/\pi$, $k_yd/\pi$.
$d=10$\,nm, and  $h=1.5\,\mu$m.
} \label{surfSlab}
\end{figure}

This property can be understood also from considering a planar waveguide
filled with a CNT array (the array axis is orthogonal to the walls of the waveguide).  The propagation constant along the waveguide is equal to
\e\l{w1}
k_{\perp}=\sqrt{\E_{zz}\left[k^2-(m\pi/2h)^2\right]},\f
where $m$ is a positive integer and $h$ is the height of the waveguide \cite{SpatD}.
If $\E_{zz}<0$, backward-wave propagation is allowed
when $k<m\pi/2h$ and forbidden for $k>m\pi/2h$.

Fig.~\ref{surfSlab} illustrates dispersion properties of three modes, corresponding to
$m=1,2$ and $m=3$. There are three  embedded conical surfaces. The internal cone corresponds to $m=1$,
and the external one to $m=3$.
One can see that backward waves propagate in the slab in a very wide
frequency range.
Their properties are quite isotropic
in the $xy$-plane due to a very small period of the CNT lattice.
It follows from formula \r{w1} that
there is no low-frequency cutoff for any mode, which can
propagate at low frequencies with very large transversal wavenumbers $k_{\perp}$.
The effective medium model becomes unapplicable if $k_{\perp}d>0.15\pi$ (compare with results of \cite{IgorPRB}).

To suppress the spatial dispersion we do not need any loadings because CNTs possess very
high kinetic inductance which ensures that
$k_z^2/n^2\ll k^2$. 
Comparing to the microwave backward-wave structures based on loaded
transmission lines and mushroom layers we note that in CNT arrays
there is no parasitic inductance between the cells and parasitic
capacitance between the tube ends and the ground plane is very small
due to the small cross-section area of the tubes. This leads to a
dramatically increased bandwidth of backward-wave propagation and gives  the ground to consider CNT arrays as 
{\bf perfect backward-wave
metamaterials}. 

For any optical material at terahertz and infrared ranges one of the
most important issues is the level of losses. In CNTs it is
determined by the relaxation time $\tau$ at the frequencies below
optical transitions. At low frequencies, including values close to a few
terahertz, $\tau$ at room temperature usually  is estimated as $3\times 10^{-12}$\,s
\cite{Slepyan,Hanson}. Such a relaxation time gives the ratio Im$(k_{\perp})$/Re$(k_{\perp})\simeq 10^{-3}$ at frequencies around 50~THz.
However, in other sources this value is taken to be $10^{-13}$\,s
\cite{achiral}, which corresponds to considerably higher losses, 
namely, Im$(k_{\perp})$/Re$(k_{\perp})\simeq 0.025$ in the same frequency range.

\begin{figure}
 \centering
\subfigure[]{\epsfig{file=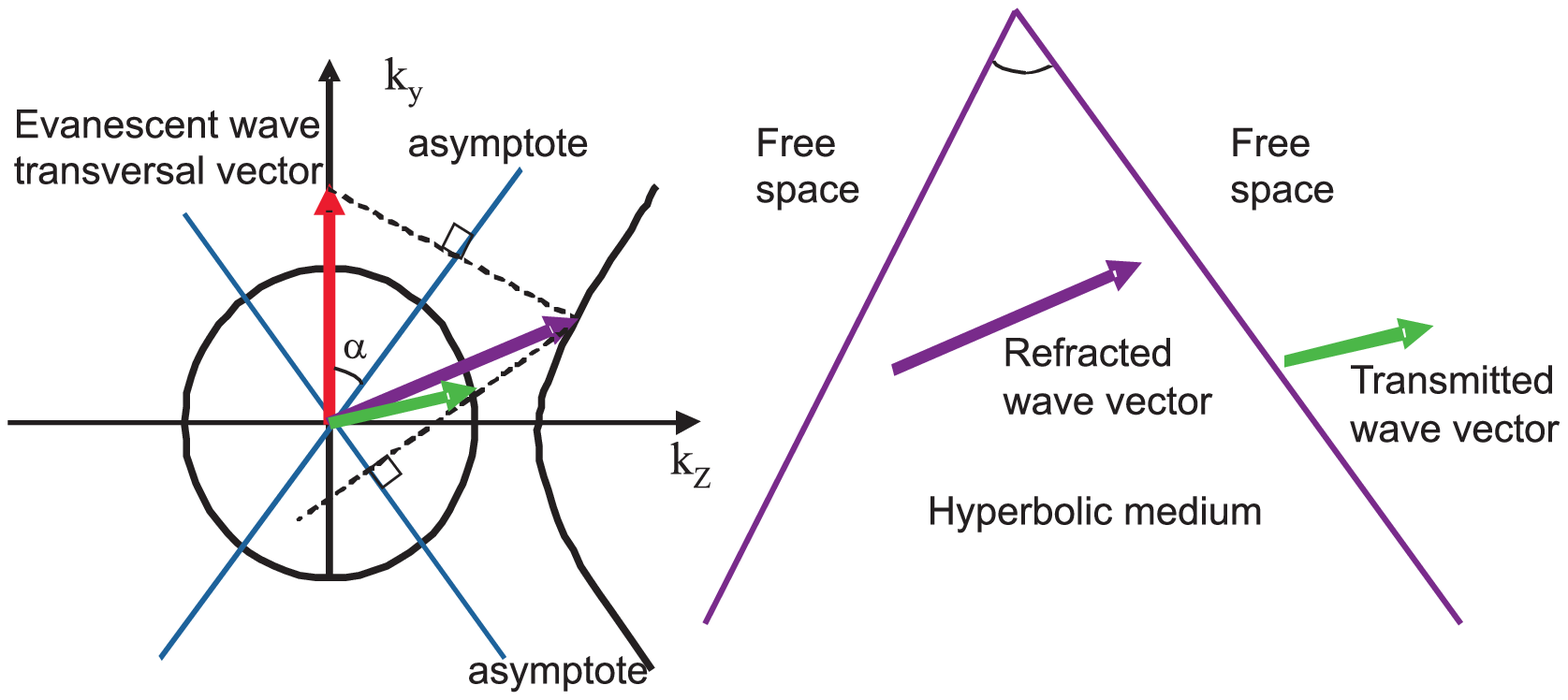,width=0.37\textwidth}}
\subfigure[]{\epsfig{file=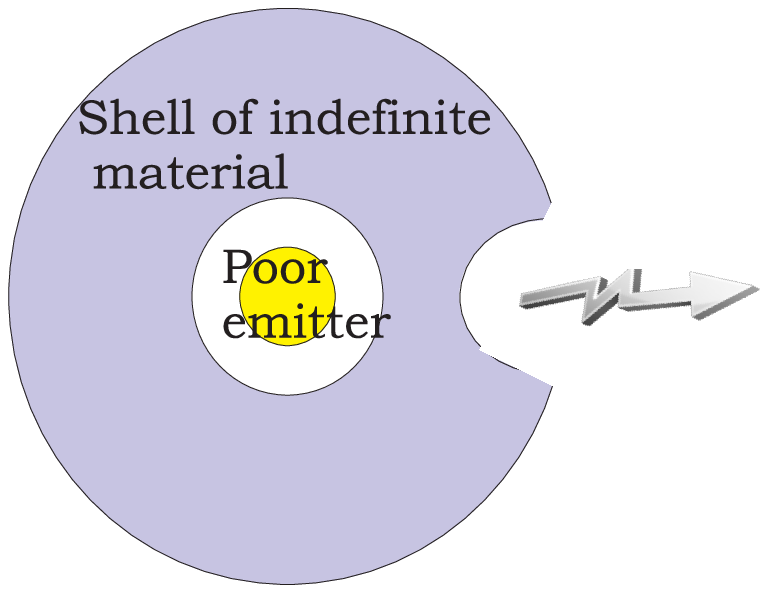,width=0.2\textwidth}}
\caption{(color online). (a) -- Conversion of evanescent waves with arbitrary
spatial frequencies into propagating ones by an indefinite wedge
with the wedge angle $2\alpha\approx \pi/2$ and horizontal optical
axis. (b) -- Enhanced emission of a poor emitter from a cut in a
shell of indefinite medium with radial directions of the local optical
axes.}
 \label{fig1}
 \end{figure}

Here we point out one potential application of indefinite
dielectric materials based on the conversion of incident evanescent waves
into propagating transmitted waves by samples of an indefinite
medium. Such a conversion is possible up to very high, in the
limiting case infinite spatial frequencies. The idea is illustrated
by Fig. \ref{fig1}.

Consider a prism with the wedge angle $2\alpha\approx \pi/2$ filled
with an indefinite medium with components of the permittivity
tensor $\varepsilon_{zz}<0$,
$\varepsilon_{yy}=\varepsilon_{xx}>0$. For a wedge made of a
CNT array this would correspond to the $z$-oriented tubes. Let an
evanescent wave attenuating along $z$ and having transversal wave
vector component $k_y>k\equiv \omega\sqrt{\varepsilon_0\mu_0} $
impinges on the left side of the prism. In the isofrequency diagram
$(k_z-k_y)$ which is depicted in Fig.~\ref{fig1} (a) the wave
vector of the refracted wave selects the point at the medium
isofrequency contour so that the component of the refracted wave
vector which is parallel to the first asymptote of the hyperbolic
isofrequency contour preserves. On the second side of the prism
the component of the refracted wave vector which is parallel to
the second asymptote of the hyperbola preserves. This holds
because both sides are tilted with respect to axis $y$ under the
same angle $\alpha$ as the asymptotes of the hyperbolic
isofrequency contour. One can easily derive for this geometry that
all incident waves with spatial frequencies $k_y<k/|\cos\alpha\cos
2\alpha|$ transmit into propagating ones ($k_y<k$) behind its
second interface. If $\alpha=\pi/4$, all evanescent waves are
transformed into propagating ones.
The isofrequency contour with $\alpha=\pi/4$ holds for
medium of parallel CNT at $k=k_p/\sqrt{2}$. The peculiarity of the
suggested structure compared to the well known Kretschmann
dielectric prism widely used in optics for coupling evanescent
optical devices (waveguides and resonators) to free space is the
possibility to convert evanescent waves with arbitrary spatial
frequencies at a certain frequency. In the Kretschmann method the
maximal spatial frequency where the conversion of the evanescent
wave into propagating wave (or reverse) is restricted by
the permittivity of the material.

The suggested effect can be practically used for the enhancement
of emission from poor emitters, i.e. light sources which low
radiation but high level of the electromagnetic energy stored in
them.
The poor emitter can be placed in a
spherical shell of an indefinite material with space-variable
optical axis. Namely, the local optical axis is directed radially
(it can be implemented as an array of radially oriented CNT). If
the shell is spherical, the evanescent waves convert on its
internal surface to propagating waves and totally internally
reflect on the outer interface.
Now we make a cut in the shell
as it is shown in Fig. \ref{fig1} (b). In this part of the shell
the inner and outer boundaries of the indefinite medium are
strongly not parallel and one can obtain the radiation
dramatically enhanced compared to the original radiation of the
emitter.

We have shown that arrays of single-wall metallic carbon
nanotubes behave as indefinite media. Such properties are provided by a very high kinetic inductance of thin carbon nanotubes. Arrays of finite-length CNTs
support propagation of
backward waves, which are characterized by
low levels of losses in the terahertz and mid-infrared ranges.
The finite-thickness  carbon nanotube
array can be considered as the perfect isotropic backward-wave metamaterial.
We have theoretically demonstrated that these materials can be used
 as transformers of full spectrum of evanescent
electromagnetic waves into propagating ones, and as devices for
emission enhancement for subwavelength-sized radiators.

This work has been partially funded by the Academy of
Finland and Nokia through the Center-of-Excellence program.


\begin{thebibliography}{9}

\bibitem{Ritchie} 
R. H. Ritchie,
Phys. Rev. {\bf 106}, 874 (1957).

\bibitem{Kretschmann} 
E. Kretschmann, Optics Communications {\bf 5}, 331 (1972).

\bibitem{Hecht} 
L., Novotny, B. Hecht, {\itshape Principles of Nanooptics} (Cambridge
University Press, New York, 2006).

\bibitem{Engheta} 
S. Bozhevolnyi, {\itshape Plasmonic nanoguides and circuits} (Pan Stanford
Publishing, Singapore, 2009).

\bibitem{Smith} 
D. R. Smith, D. Schurig, Phys. Rev. Lett. \textbf{90}, 077405 (2003).

\bibitem{Pendry} 
J. B. Pendry, Phys. Rev. Lett. \textbf{ 85}, 3966 (2000).

\bibitem{Zhang} 
X. Zhang, Z. Liu,  Nature Materials \textbf{7}, 435 (2008).


\bibitem{Smith3} 
Y. Liu, G. Bartal, X. Zhang, Optics Express \textbf{16}, 15439 (2008).

\bibitem{Brown1} 
J. Brown, Proc. IEE \textbf{100}, 51 (1953).

\bibitem{Brown2} 
J. Brown, W. Jackson, Proc. IEE  \textbf{102B}, 11 (1955).

\bibitem{SpatD} 
P.A. Belov, R. Marques, S.I. Maslovski, I.S. Nefedov,
 M. Silveirinha, C.R.~Simovski and S.A.~Tretyakov,
Phys.~Rev.~B  {\bf 67}, 113103 (2003).


\bibitem{Hudl} 
M. Hudli\v{c}ka, J. Macha\v{c}, I. Nefedov, Progress In
Electromagnetics Research, PIER {\bf 65}, 233 (2006).

\bibitem{Demetr} 
A. Demetriadou, J. B.  Pendry, J. Phys.: Condens. Matter {\bf 20}, 295222 (2008).

\bibitem{Grib} 
A. B. Yakovlev {\itshape et al.},
 M. G. Silveirinha, O. Luukkonen, C. R.  Simovski, I. S. Nefedov, S. A.  Tretyakov,
 IEEE Trans. on Microw. Theory and
Tech. {\bf 57}, 2700 (2009).

\bibitem{stas} 
S. Maslovski, M. Silveirinha, Phys. Rev. B {\bf 80}, 245101 (2009).


\bibitem{Lutt} 
P. J. Burke, IEEE Trans. on Nanotechnology {\bf 3}, 129 (2002).

\bibitem{ShFan} 
S. Fan {\itshape et al.},  Science {\bf 283}, 512 (1999).

\bibitem{YLin} 
Y. Lin, F. Lu, Y. Tu, Z. Ren,  Nano Letters {\bf 4}, 191 (2004).

\bibitem{Wang} 
Y. Wang {\itshape et al.}, Appl. Phys. Lett. {\bf 85}, 2607 (2004).

\bibitem{DresselNat} 
M. S. Dresselhaus, Nature {\bf 432}, 959 (2004).

\bibitem{WonBong} 
W. B. Choi {\itshape et al.}, Nanotechnology {\bf 15}, S512 (2004).

\bibitem{Kusunoki} 
M. Kusunoki {\itshape et al.}, Chem. Phys. Lett. {\bf 366}, 458 (2002).

\bibitem{modified} 
A. M. Nemilentsau {\itshape et al.}, Phys. Rev. B {\bf 82}, 235424 (2010).


\bibitem{Slepyan} 
G.Y. Slepyan, S. A. Maksimenko, A. Lakhtakia, O. Yevtushenko and A. V. Gusakov, Phys. Rev. B {\bf 60}, 17136 (1999).

\bibitem{BurkeNT} 
P.J. Burke, S. Li, and Z. Yu,
IEEE Trans. on Nanotechnology {\bf 5}, 314 (2006).

\bibitem{Landau} 
L.D. Landau and E.M. Lifshitz, {\itshape Electrodynamics of
Continuous Media (Course of Theoretical Physics, Volume 8)}
(Butterworth-Heinemann; 2 edition, 1984).



\bibitem{IgorPRB} 
I. S. Nefedov, Phys. Rev. B {\bf 82}, 155423 (2010).

\bibitem{Elefther} 
G. V. Eleftheriades, K. G. Balmain, eds. \emph{Negative-Refraction Metamaterials: Fundamental Principles
and Applications} (Hoboken, NJ: J. Wiley and Sons, 2005).

\bibitem{Caloz1} 
C. Caloz, T. Itoh,  \emph{Electromagnetic metamaterials:
transmission line theory and microwave applications} (New York:
J. Wiley and Sons, 2006).

\bibitem{Olli} 
O. Luukkonen {\itshape et al.},
P. Alitalo, F. Costa, C. Simovski, A. Monorchio, S. Tretyakov
 Appl. Phys. Lett. {\bf  96}, 081501 (2010).

\bibitem{Hanson} 
G. V. Hanson, IEEE Trans. on Ant. and
Prop. {\bf 53}, 3426 (2005).

\bibitem{achiral} 
G. Ya. Slepyan, M. V. Shuba, S. A. Maksimenko, and A. Lakhtakia, Phys. Rev. B {\bf 73},
195416 (2006).




\end{thebibliography}
\end{document}